\documentclass[11pt,letterpaper]{article}
\usepackage{float}
\usepackage{multirow}
\usepackage{ucs,amssymb,amsmath}
\usepackage[utf8]{inputenc}
\usepackage{fontenc}
\usepackage{graphicx}
\usepackage[pdftex,colorlinks]{hyperref,ragged2e}
\usepackage[super,sort&compress,numbers]{natbib}

\usepackage[margin=1in]{geometry}
\usepackage{setspace}

\newif\ifMarked
\Markedtrue 

\ifMarked

\else

\fi

\title{Assessment of P-value variability in the current replicability crisis}
% ...
\begin{document}
\maketitle
\thispagestyle{empty}
\begin{flushleft}
\author{Olga Vsevolozhskaya, Ph.D.,$^{1}$ Gabriel Ruiz,$^{2}$ Dmitri Zaykin, Ph.D.$^{3}$\\
\vskip 2ex
   $^1$Biostatistics Department, University of Kentucky, Lexington, KY, USA\\
   $^2$The Summer Internship Program at the National Institute of Environmental Health Sciences, Research Triangle Park, NC, USA \\
   $^3$Biostatistics and Computational Biology, National Institute of Environmental Health Sciences, National Institutes of Health, Research Triangle Park, NC, USA}
\vskip 40ex
Correspondence: Dmitri V. Zaykin, Senior Investigator at the Biostatistics and Computational Biology Branch, National Institute of Environmental Health Sciences, National Institutes of Health, P.O. Box 12233, Research Triangle Park, NC 27709, USA. Tel.: +1 (919) 541-0096 ; Fax: +1  (919) 541-4311. Email address: dmitri.zaykin@nih.gov
\end{flushleft}

\clearpage
\setcounter{page}{1}
\section*{Abstract}
Increased availability of data and accessibility of computational tools in recent years have created unprecedented opportunities for scientific research driven by statistical analysis. Inherent limitations of statistics impose constrains on reliability of conclusions drawn from data but misuse of statistical methods is a growing concern. Significance, hypothesis testing and the accompanying P-values are being scrutinized as representing most widely applied and abused practices. One line of critique is that P-values are inherently unfit to fulfill their ostensible role as measures of scientific hypothesis's credibility. It has  also been suggested that while P-values may have their role as summary measures of effect, researchers underappreciate the degree of randomness in the P-value. High variability of P-values would suggest that having obtained a small P-value in one study, one is, nevertheless, likely to obtain a much larger P-value in a similarly powered replication study. Thus, ``replicability of P-value'' is itself questionable. To characterize P-value variability one can use prediction intervals whose endpoints reflect the likely spread of P-values that could have been obtained by a replication study. Unfortunately, the intervals currently in use, the P-intervals, are based on unrealistic implicit assumptions. Namely, P-intervals are constructed with the assumptions that imply substantial chances of encountering large values of effect size in an observational study, which leads to bias. The long-run coverage property provided by P-intervals is similar in interpretation to the coverage provided by the classical confidence intervals, but the endpoints of any particular interval lack interpretation as probabilistic bounds for possible spread of future P-values that may have been obtained in replication studies.  As an alternative to P-intervals, we develop a method that gives researchers flexibility by providing them with the means to control these assumptions. Unlike endpoints of P-intervals, endpoints of our intervals are directly interpreted as probabilistic bounds for replication P-values and are resistant to selection bias contingent upon approximate prior knowledge of the effect size distribution. We showcase our approach by its application to P-values reported for five psychiatric disorders by the  Psychiatric Genomics Consortium group.
\clearpage
\section*{Introduction}
Poor replicability has been plaguing observational studies. The ``replicability crisis'' is largely statistical and while there are limits to what statistics can do, a serious concern is misapplication of statistical methods. Significance testing and P-values are often singled out as major culprits not only because these concepts are easy to misinterpret but for purported inherent flaws. It has been suggested that variability of P-values is underappreciated in the sense that when a small P-value is obtained by a given study, researchers commonly suppose that a similarly designed independent replication study is likely to yield a similarly small P-value. Great variability of replication P-values casts doubt on validity of conclusions derived by a study at hand and implies lack of confidence in possible outcomes of any follow-up studies. It has been suggested that in reality one should expect a great uncertainty in what replication P-value will be and special prediction intervals for P-values (``P-intervals'') have been employed to characterize that uncertainty.\cite{cumming2008replication, lai2012subjective, halsey2015fickle, lazzeroni2014p, lazzeroni2016solutions} 

P-intervals have been presented as an objective measure of P-value variability, as opposed to subjective judgements reported by researchers in surveys, with the conclusion that the subjective estimates are too narrow and therefore researchers tend to underestimate randomness of replication P-values.\cite{lai2012subjective} While P-intervals have been used mainly as a tool to elucidate flaws of P-values, they have also been promoted as important additions to P-values themselves in publications supportive of P-values as being useful and universal summaries of statistical tests. It has been suggested that P-intervals improve P-value interpretability, especially in large-scale genomic studies with many tests and other studies utilizing modern high-throughput technologies. \cite{lazzeroni2014p,lazzeroni2016solutions}

The distribution of P-value and thus its variability is easily characterized analytically for the basic test statistics and depends on a measure of effect size such as the value of the relative risk (RR) of disease given exposure vs. nonexposure to a pollutant. Effect sizes themselves can be thought as arising from a distribution, e.g., see Equation 11 in Kuo et al. 2015,\cite{kuo2015assessing} in which case the P-value distribution becomes a weighted average, i.e., a marginal distribution over all possible values of the effect with their respective probabilities as weights. The idea behind the P-intervals is that even without such ``prior'' knowledge on possible values of the effect size, one can take at face value the magnitude of an obtained P-value (P$_\text{obt}$). In other words, there is information about the magnitude of the effect size contained in the magnitude of P$_\text{obt}$ which can be used to make predictions about a P-value obtained in a replication study, denoted as P$_\text{rep}$.

The following quote from Cumming, 2008 \cite{cumming2008replication} gets to the heart of the matter succinctly: ``\textit{This article shows that, if an initial experiment results in two-tailed p = .05, there is an 80\% chance the one-tailed p value from a replication will fall in the interval (.00008,  .44)} [...] \textit{Remarkably, the interval -- termed a p interval -- is this wide however large the sample size.}'' An equivalent statement appears in a \textit{Nature Methods} letter by Halsey and colleagues: ``\textit{regardless of the statistical power of an experiment, if a single replicate returns a P value of 0.05, there is an 80\% chance that a repeat experiment would return a P value between 0 and 0.44.}''\cite{halsey2015fickle}  Both statements make use of a specific value, P=0.05 for which the interval is constructed and the endpoints of that interval are described explicitly as probability bounds for possible values of replication P-values. However, these P-intervals are derived as classical prediction intervals based on the normal test statistic (Z-statistic) and as such only provide long-run $\beta$\% coverage, meaning that $\beta$\% of P-intervals constructed around {\em different values} of P$_\text{obt}$ over a large number of studies would contain the respective replication P-values. It is important that by construction, the coverage is guaranteed as an average over many studies only {\em without any restrictions on the value of} P$_\text{obt}$. 

Once a P-interval is constructed for a particular P$_\text{obt}$, its endpoints cannot be interpreted as statistical bounds for a replication P-value, unless specific conditions are met regarding the likelihood of all possible values for the standardized effect size, such as the difference between two sample means over the standard deviation, $\delta$=$(\mu_1-\mu_2)/\sigma$. If one took a large collection of observational studies and had access to the actual, population values of $\delta$, then the distribution of these values would have had the variance that we denote by $s^2_0$. When a P-interval is constructed assuming the sample size $N$, its endpoints can be interpreted as probability bounds for P$_\text{rep}$ only when $N\times s^2_0$ approaches infinity. This is equivalent to the assumption that the expected value of the Z-statistic for a randomly selected study of size $N$ can be anywhere from $-\infty$ to $\infty$ with equal probability. 

How large can $N\times s^2_0$ be assumed to be realistically? Genetic epidemiology and other observational studies routinely test hypotheses that can be viewed conceptually as a comparison of two sample means. Exposure to an environmental factor or genetic effect of a locus on susceptibility to disease are examples where the presence of effect implies a difference in mean values between subjects with and without disease. In these examples, the effect size can be measured by the log of odds ratio, log(OR). Expecting the majority of effect sizes to be small and the direction of effect to be random, log(OR) can be described by a zero-centered, bell-shaped distribution. It can be shown (Methods section) that the value $\delta$ for a given value of log(OR) cannot exceed $\log(\text{OR})/ \left[2 \sqrt{2 + (1+\text{OR}) / \sqrt{\text{OR}}}\right]$. This implies that a considerable spread of $\delta$ values is unrealistic. For example, OR=4 gives the maximum possible value for $\delta$ to be about 1/3. Even a very large value OR=10 results in  max($\delta$) $\approx$ 1/2. Such large odds ratios are very rarely encountered in epidemiological studies, suggesting that realistic variance values $s^2_0$ cannot be very large. Further, as detailed in Methods section, the maximum possible value of $\delta$ for any OR cannot exceed $\approx$0.663. This bound places further restrictions on realistic and maximum possible values of the prior variance $s^2_0$, because the prior distribution has to vanish at that bound.

Genetic epidemiology studies and genome-wide association scans, in particular, routinely involve massive testing. These studies have uncovered many robustly replicating genetic variants that are predictors of susceptibility to complex diseases. It is also apparent that the vast majority of genetic variants carry effect sizes, such as measured by log(OR), that are very close to zero, and there are commonly only a handful of variants with odds ratios as large as 1.5. This implies tiny values of $s_0^2$. For example, a reported distribution of effect sizes for the bipolar disorder\cite{chen2013genome} and cancers\cite{park2010estimation} translates into the values of the order $10^{-6}$ to $10^{-5}$ for $s_0^2$ (Methods section). In this paper, we show how small values of $s_0^2$ render P-intervals unfit as a measure of P$_\text{rep}$ variability  and provide an alternative solution based on a Mixture Bayesian approach, which is not constrained to the conjugate model only and provides a researcher with the flexibility to specify any desired prior effect size distribution. Our results reveal immunity of the Mixture Bayes-based intervals to selection bias and multiple testing phenomena. We contrast the performance of the traditional P-intervals relative to the Bayesian-based prediction intervals using results from the Psychiatric Genomics Consortium (PGS)\cite{lazzeroni2014p} and conclude with a discussion of the implications of our findings.

\section*{Methods}
\subsection*{Prediction intervals}
The P-interval can be obtained as a classical prediction interval for the normally distributed test statistics, (Z-statistics). The prediction distribution for the statistic $Z_\text{rep}$ that relates to the one-sided P-value as $P_\text{rep} = 1 - \Phi(Z_\text{rep})$ is the normal, $\Phi(z_\text{obt}, 2)$, where $\Phi(\cdot)$ is the standard normal cumulative distribution function (CDF). This distribution does not depend on the actual mean of $Z$, which is $\sqrt{N}\times \delta$. The reason for that becomes apparent when the P-interval is derived as a Bayesian prediction interval. For a normally distributed Z-statistic, $Z \sim N(\mu, 1)$, assume the conjugate model, that is, $\mu \sim \sqrt{N} \times \Phi(m_0, s^2_0)$. Then the posterior distribution for the mean of $Z_\text{obt}$ is normal $\Phi(\theta, s^2)$, where
\begin{eqnarray}
  \theta \mid Z_\text{obt} &=& \frac{\frac{m_0}{\sqrt{N} s^2_0} + Z_\text{obt}}{s^2}  \\
  s^2 \mid Z_\text{obt} &=& \left[ \frac{1}{N s^2_0} + 1 \right]^{-1}, 
\end{eqnarray}
and the prediction distribution for $Z_\text{rep}$ is $\Phi(\theta, 1+s^2)$. Thus, the P-interval based on the distribution $\Phi(z_\text{obt}, 2)$ is a Bayesian interval that implicitly assumes that $N \times s^2_0 \rightarrow \infty$. We refer to the resulting intervals as the Conjugate Bayes intervals. The endpoints of these intervals are given by
\begin{eqnarray}
  && \theta \pm z_{(1-\alpha/2)} \sqrt{s^2+1} \\
  &&  \equiv Z_\text{obt} \frac{\sigma_0^2}{1+\sigma_0^2}  + \frac{m_0}{1+\sigma_0^2}
   \pm z_{(1-\alpha/2)} \sqrt{1+\frac{\sigma_0^2}{1+\sigma_0^2}},
\end{eqnarray} 
where $z_{(1-\alpha/2)}$ is the $1-\alpha/2$ quantile of the standard normal distribution and $\sigma_0^2=N s^2_0$. The conjugate model is restrictive in that a specific prior distribution has to be assumed, which may not provide an adequate representation of external knowledge about the effect size distribution. It also limits construction of the intervals to P-values derived from statistics for which there are known conjugate priors. Here, we introduce a much more flexible approach, the Mixture Bayes, without these restrictions. The Mixture Bayes intervals can be constructed for P-values derived from statistics whose distribution is governed by a parameter $\gamma$ that captures deviation from the usual point null hypothesis, $H_0$, and has the form $\sqrt{N} \times \delta$ or its square, $N \times \delta^{2}$. This includes normal, chi-squared, Student's $t$ and F-statistics. We partition the prior distribution of $\gamma$ into a finite mixture of values $\delta_1, \delta_2, \dots, \delta_B$ with the corresponding prior probabilities, $\Pr(\delta_i)$. As an example, let P-value be derived from an F-test for comparison of two sample means, with the corresponding sample sizes $n_1$ and $n_2$. Let $N = 1/(1/n_1 + 1/n_2)$.
For $i$-th prior value of effect, a statistic based on sampling values of
$T = (\bar{X}_1 - \bar{X}_2)^2 / \hat{\sigma}^2$ has a noncentral F distribution, with the noncentrality 
\begin{eqnarray}
   \gamma_i = N \left[ (\mu_1 - \mu_2) / \sigma \right]^2_i = N \delta^2_i,
\end{eqnarray}
and the degrees of freedom $\text{df}_1=1, \text{df}_2=n_1+n_2-2$:
\begin{eqnarray}
   T \sim f(T = t \mid \gamma_i, \text{df}_1, \text{df}_2),
\end{eqnarray}
where $f$ is the density of the noncentral F-distribution. The posterior distribution is a mixture,
\begin{eqnarray}
   \Pr(\delta^2_j \mid T=t) = \frac{\Pr(\delta^2_j) f(T = t \mid \gamma_j, \text{df}_1, \text{df}_2)}{\sum_{i=1}^B \Pr(\delta^2_i) f(T = t \mid \gamma_i, \text{df}_1, \text{df}_2)},
\end{eqnarray}
with the posterior mean
\begin{eqnarray}
   \theta = \sum_{i=1}^B \delta^2_i \,\,  \Pr(\delta^2_i \mid \text{P-value}).
\end{eqnarray}
Next, we obtain the CDF of the prediction distribution for the replication statistic, $T_\text{rep}$, as
\begin{eqnarray}
     F_p(x) &=& \sum^B_j \Pr(\delta^2_j \mid T_\text{obt} ) \int^x_0 f(T_\text{rep} \mid \gamma_j, \text{df}_1, \text{df}_2) d T_\text{rep} \nonumber \\
  &=& \sum^B_j \Pr(\delta^2_j \mid T_\text{obt} ) F(T_\text{rep}=x \mid \gamma_j, \text{df}_1, \text{df}_2).
\end{eqnarray}
Then, the Mixture Bayes interval endpoints are derived from the quantiles of this CDF, that are given by $F^{-1}_p(x)$.

\subsection*{Prior variance for the standardized logarithm of the odds ratio}
Genetic epidemiology and other observational studies routinely test hypotheses conceptually related to a comparison of two sample means.  Effect size is often measured by the log of odds ratio, which can be related to the difference in means (that become frequencies, $p_1$ and $p_2$, in the case of binary variables) as $p_1-p_2 \approx \log(\text{OR}) \, \tilde{p}(1-\tilde{p})$, where $\tilde{p}$ is the pooled frequency. Distribution of P-values for commonly used test statistics depends on the product of the sample size, ($N$ or $\sqrt{N}$), and a measure of effect size, $\mu$, scaled by the variance $\sigma^2$ (or $\sigma$), i.e. $\delta=\mu / \sigma$. For example, when the outcome is a case/control classification and the predictor is also binary, the standardized effect size can be expressed in terms of the correlation ($R$) times the sample size as follows:
\begin{eqnarray}
   \gamma &=&  \sqrt{N}\times \frac{\mu}{\sigma} = \sqrt{N} \times \delta = \sqrt{N} \times R \nonumber \\
  &=& \sqrt{N} \times \frac{p_1-p_2}{\sqrt{\tilde{p}(1-\tilde{p}) \left[v(1-v)\right]^{-1}}}, \\
\end{eqnarray}
where $v$ is the proportion of cases in the sample. In terms of the logarithm of the odds ratio, OR,
\begin{eqnarray}
  \gamma &=& \sqrt{N}\times \delta = \sqrt{N} \times \frac{\log(\text{OR})}{\sqrt{\frac{1}{v} \frac{1}{p_1(1-p_1)} + \frac{1}{1-v} \frac{1}{p_2(1-p_2)}}} \label{logor} \\
     &\approx& \sqrt{N} \times \frac{\log(\text{OR})}{ \sqrt{\left[\tilde{p}(1-\tilde{p}) v(1-v)\right]^{-1} }}.
\end{eqnarray}
For a given value of OR, the standardized effect size $\delta$  cannot exceed the value $\delta_{\text{max}}(\text{OR})$ that we obtained by maximizing the right hand side of Equ. (\ref{logor}) as:
\begin{eqnarray}
 \delta_{\text{max}}(\text{OR})  &=& \frac{\ln(\text{OR})}
  {2 \sqrt{2 + \frac{1+\text{OR}}{\sqrt{\text{OR}}}}}. \label{maxor}
\end{eqnarray}
Let $F^{-1}(\cdot \mid \mu_0, s_0^2)$  denote the inverse CDF of the conjugate prior distribution. Writing $\Pr(\text{OR} \ge x)=\beta$ and assuming a symmetric distribution of the effect size around zero, i.e., $m_0=0$, we can relate the value $\delta_{\text{max}}$ to the prior variance of the conjugate model in the following way:
\begin{eqnarray}
    \delta_{\text{max}}(\text{OR}) &=&\sqrt{s_0^2} \, F^{-1}(1-\beta \mid 0,1). \nonumber
\end{eqnarray}
The maximum spread for the conjugate prior distribution is therefore obtained when its variance is equal to
\begin{eqnarray}
   s_0^2 = \left[ \frac{\delta_{\text{max}}(\text{OR})}{F^{-1}(1-\beta \mid 0,1)} \right]^2.
 \label{uvalue}
\end{eqnarray}
It should be noted that Equ. \ref{maxor} gives the maximum $\delta$ value for a given value of OR, however it is not monotone in OR. The maximum possible value of $\delta_{\text{max}}(\text{OR})$ can be found to be at OR $\approx$ 121.35. Curiously, this value of OR implies $\delta_{\text{max}}(\text{OR})$ value equal to the Laplace Limit constant, 0.662743...

Park et al.\cite{park2010estimation} reported distribution of effect sizes for breast, prostate and colorectal (BPC) cancers in terms of a table, giving the numbers of different loci ($L_i$) with the corresponding values of OR$_i$. Using the same approach, Chen et al.\cite{chen2013genome} provided the effect size distribution for the bipolar disorder (BP) risk loci. Assuming the total number of independent variants to be $M$=300 000, proportions of associated loci are $w_i=L_i/M$. We assumed the average OR among non-associated loci to be 1.005 (or its inverse for the negative part of the log(OR) distribution). The variance $s_0^2$ was calculated as $\sum_i w_i (\gamma_i/\sqrt{2N} - m_w)^2$, where $m_w = \sum_i w_i \gamma_i /\sqrt{2N}$, and gave the value $\approx 5\times10^{-6}$ for both cancer and the BP disorder risk loci. Thus, $N$ needs to be about 50 000 for $s_0^2 \times N$ to reach 1/2.

\section*{Results}
\noindent Table \ref{tab:Zselection} summarizes the empirical coverage probabilities of the 80\% prediction intervals for a standardized effect size $\sqrt{N}\times \delta$ under different types of  P-value selection (simulation study set up is detailed in Supplementary Information). The observed P-value was based on a two-sample Z-test and was thresholded according to the following selection rules: (i) no selection, i.e., a prediction interval is constructed for a randomly observed P-value; (ii) selection of P-values around a value, e.g., $P \approx 0.05$, i.e., prediction intervals are constructed only for P-values that were close to the 5\% significance level; (iii) selection of P-values that are smaller than a threshold, e.g., $P <0.05$. The empirical coverage was calculated based on 50 000 simulations using three different methods: (a) a conjugate Bayesian model assuming normal prior distribution for the observed value of a test statistic, $Z_\text{obt} \sim \Phi(0, \sigma_0^2)$, where $\sigma_0^2=s_0^2 \times N$; (b) our Mixture Bayes approach with the same prior as for the conjugate model; and (c) the original P-interval proposed by Cumming.\cite{cumming2008replication} 

Mixture Bayes intervals were included in these simulations to check how well they approximate a continuous prior distribution assumed by the conjugate intervals. We used mixture components with the length $\sigma_0/8$ for every component and truncated the normal prior at 10$^{-6}$ and $1-10^{-6}$ quantiles. This provided us with sufficient accuracy and resulted in the number of mixture components, $B$, equal to 76 for all values of $\sigma_0^2$.

Table \ref{tab:Zselection} clearly indicates that all three construction methods have the correct coverage ($\sim$80\%) if a prediction interval is calculated for a randomly observed P-value $\in \left[0, 1\right]$. However, selection and small prior variance both impair performance of P-intervals. For instance, if an interval is constructed for a P-value $< 0.001$ and $\sigma_0^2 =0.25$, the coverage of the traditional non-Bayesian P-interval may be as low as 17\%. Even for large values of prior $\sigma^2_0$, the P-interval has poor coverage when constructed for P-values around genome-wide significance levels (e.g., $\text{P-value} < 1.5 \times 10^{-7}$).

Similar conclusions regarding the coverage can be drawn if a prediction interval is constructed for the most significant P-value out of $L$ tests (Table \ref{tab:ZminP}). Specifically, both Bayesian methods have the correct coverage and are immune to the selection bias. The non-Bayesian P-interval approach, however, once again performs poorly if the prior variance is small. Additionally, as the number of tests increases, out of which a minimum P-value is selected, the P-interval coverage is becoming increasingly off the 80\% mark.

We next explored the effect of prior variance mis-specification on the coverage of the Bayesian-type prediction interval when it is constructed for the most significant result out of $L$ tests. Two scenarios were considered: under-specification ($\sigma_0^2/2$) and over-specification ($2 \sigma_0^2)$ of the prior variance $\sigma_0^2 = N s_0^2$. The results are summarized in Table \ref{Tab:bias} and indicate that in terms of the coverage it is safer to over-specify values of the prior variance than to under-specify them. The conjugate model with $m_0=0$ gives the intervals in the following form
\begin{eqnarray}
    Z_\text{obt} \frac{\sigma_0^2}{1+\sigma_0^2}
  \pm z_{(1-\alpha/2)} \sqrt{1+\frac{\sigma_0^2}{1+\sigma_0^2}},
\end{eqnarray} 
indicating that $\sigma_0^2$ values that are too small pull the interval mean excessively toward zero while at the same time reducing its proper length.

Unlike the regular Bayesian model, our Mixture Bayes approach is not limited to conjugate priors and prediction intervals can be constructed for any P-value stemming from statistics other than the normal Z-test. Additionally, the Mixture Bayes approach allows the use of any prior distribution and enjoys the same coverage properties as the conjugate-Bayes prediction intervals, that is, resistance to multiple testing and selection bias (Supplementary Tables \ref{tab:si1} and \ref{tab:si2}). 

To illustrate the interpretation of the prediction intervals, we replicated part of Table 1 in Lazzeroni et al.,\cite{lazzeroni2014p} who considered recent findings from the Psychiatric Genomics Consortium  (PGC) for attention deficit-hyperactivity disorder (ADHD), autism spectrum disorder (ASD), bipolar disorder (BPD), major depressive disorder (MDD) and schizophrenia. The consortium reported four single nucleotyde polymorphisms (SNPs) associated with these psychiatric disorders but, for illustrative purposes, we constructed prediction intervals only for a single SNP, rs2535629. We used four different methods to calculate prediction intervals: (i) the conjugate Bayesian model with the estimated prior variance, $s^2_0$, based on the results from Chen et al.\cite{chen2013genome} (see Methods section); (ii) Mixture Bayes approximation to this continuous conjugate normal prior, using the same variance, $s^2_0$; (iii) Mixture Bayes approach with the BP effect size distribution reported in Chen et al. as a prior (without assuming the conjugate model); and (iv) prediction intervals suggested by Lazzeroni et al. (which are equivalent to Cumming's P-intervals for one-sided P-values). %

Table \ref{tab:bp} summarizes the results.  Due to our usage of intervals for one-sided P-values in the last column of our Table \ref{tab:bp}, there is a discrepancy with the prediction intervals given in Table 1 of Lazzeroni et al. For instance, for the ADHD disorder, Lazzeroni and colleagues provided prediction interval bounds of (0.00, 4.29). These values can be easily recovered from Table \ref{tab:bp} by subtracting logarithm based ten of two from both prediction interval bounds (e.g., 4.29 = 4.59 $- \log_{10}(2)$).

For all psychiatric disorders, lower bounds of the 95\% prediction intervals for P-values based on the approach suggested by Lazzeroni et al.\cite{lazzeroni2014p} are smaller (or larger in terms of the -$\log_{10}(\text{P-value})$ interval) than the ones from the Bayesian-based methods. For instance, Lazzeroni and colleagues concluded that in a similarly powered replication of the original PGC design, a P-value for an association between rs2535629 and ADHD could be as low as $2.5 \times 10^{-5}$, given the observed P-value of 0.1005. Our interval results portray a less optimistic picture with the P-value lower bound for ADHD equal to 0.02.
Similar observations hold for psychiatric disorders with significant observed P-values. For example, in Lazzeroni et al. the bipolar disorder is concluded to be likely to yield a P-value $<10^{-8}$ for rs2535629, reaching genome-wide significance at a replication study. Mixture Bayes prediction interval based on the reported effect size distribution\cite{chen2013genome} suggests a higher lower bound for the BP replication P-value of $6.9 \times 10^{-7}$. The difference highlights the implicit prior assumption built into the P-intervals that large effect sizes are as likely as small a priori.

Nonetheless, similar to conclusions in Lazzeroni et al., the association of rs2535629 with BP appears to be a promising signal. Also, similarly to the conclusions in Lazzeroni et al., the combined study of all psychiatric disorders is predicted to perform better than replication studies of individual phenotypes. 

While it is expected that different diseases would have different effect size distributions, we wanted to check the robustness of our results to prior mis-specification and utilized available effect size distribution given in Park et al.\cite{park2010estimation} for cancers. This assumes that the effect size distribution in terms of odds ratios has common main features for different complex diseases, namely, that it is L-shaped  with the majority of effect sizes that can be attributed to individual SNPs being very small, and that the frequency of relatively common variants with increasingly large values of OR quickly dropping to zero for OR as large as about 3. The modified intervals are reported in Table \ref{tab:cancer}. While Mixture Bayes intervals become somewhat different from those derived using the effect size distribution for BP, their bounds are much more similar to each other than to the bounds of P-intervals.

\section*{Discussion}
It can be argued that regardless of the degree of their variability, P-values are poorly suited for what they are used for in practice. Researchers want to know whether a statistic used for summarizing their data supports their scientific hypothesis and to what degree. P-values in general do not reflect uncertainty about a hypothesis. This point and other misconceptions have been recently reviewed in a statement on statistical significance and P-values by the American Statistical Association.\cite{wasserstein2016asa} Here, we focused specifically on variability of P-values in replication studies. We examined implicit prior assumptions of previously suggested methods and detailed how these assumptions can be explicitly stated in terms of the distribution of the effect size. As an intermediate step of our approach, fully Bayesian posterior distributions for standardized parameters, such as $(\mu_1-\mu_2)/\sigma$, are readily extractable from P-values that originate from many basic and widely used test statistics, including the normal Z-statistics, Student's t-test statistics, chi-square and F-statistics.

Our results show that while P-intervals are derived without the explicit assumption that all effect sizes are equally likely, such a ``flat'' prior is assumed implicitly and leads to bias when in reality the effect size distribution is modeled more realistically, allowing only a small chance to encounter a large effect size and supposing that the majority of effect sizes would be small. Many observational studies seeking associations of health outcomes with environmental exposures and genetic predictors can be viewed conceptually as a comparison of two sample means, $\delta=\mu_1-\mu_2$. Presence of a true association in such studies implies a certain difference in mean values of exposure between subjects with and without disease. In such examples, the prior variance reflects the prior spread of the mean of a test statistic, which usually can be related to the spread of the standardized mean difference. The prior spread in units of standard deviation cannot be very large, especially in the fields of observational sciences, that are currently at the focus of the replicability crisis. For example, assuming that effect sizes with the odds ratio (OR) greater than three are relatively rare (1\% occurrence rate), the prior variance for $\ln(\text{OR})/\sqrt{\text{Var}(\ln(\text{OR}))}$ is about 0.01 at its largest possible value (Equ. \ref{uvalue}) and would typically be smaller.

As we show here, setting the mean of the prediction distribution to $z_\text{obt}$ in the construction of P-intervals is equivalent to assuming that any possible value of $\sqrt{N} \times \delta$ is equally likely. This covers the entire real line $(-\infty, \infty)$ and can be an extremely strong assumption to make in many epidemiological contexts. As a consequence, interval statements obtained with P-intervals become invalid when applied to a particular value or a range of P-values (e.g. 0.049 $ < P_\text{obt} < $ 0.051) and result in biased intervals. The $(1-\alpha)$ nominal coverage of P-intervals can be Bonferroni-adjusted\cite{lazzeroni2014p} for $L$ tests as $(1-\alpha/L)$ and while that procedure can restore the long-run coverage property, the endpoints of such intervals would still lack interpretation as probability bounds for a replication P-value. On the other hand, Bayesian prediction intervals that acknowledge the actual variability in the possible values of the effect size do depend on the sample size and have correct coverage regardless of whether a selection of P-values is present. Reanalysis of the intervals reported by Lazzeroni and colleagues\cite{lazzeroni2014p} shows that P-intervals can be substantially different from Bayesian prediction intervals, even when sample sizes are very large (Table \ref{tab:bp}).  These results also reflect discrepancies obtained with the direct, ``as is'' usage of the estimated prior distribution in the Mixture Bayes approach and an attempt to approximate this distribution by the conjugate prior with the same variance. Endpoints of the conjugate intervals on the log scale are comparatively shorter and highlight lack of flexibility inherent in the conjugate approximation to the prior: allowance for a large fraction of effect sizes to be close to zero makes the tails of the conjugate distribution too thin. The estimated prior distribution used by the Mixture Bayes approach is more fat-tailed and is also asymmetric due to a high proportion of minor alleles that carry effects of the positive sign. 

Bayesian prediction intervals require informed input about various values of the effect size and their respective frequencies. This is not impossible. We know, for example, that in genetic association studies, the squared effect size distribution is L-shaped, where the majority of genetic effects across the genome are tiny and only few are large. We should contrast such a prior distribution, even if specified only approximately, with a largely unrealistic assumption implicit in P-intervals. When the assumed prior distribution does in fact follow reality, Bayesian prediction intervals enjoy the property of being resistant to the winner's curse. One can select P-values in any range and obtain unbiased intervals or select the minimum P-value from an experiment with however many tests: the resulting interval would still be unbiased without the need of a multiple-testing adjustment to its coverage level. P-intervals are a special case of our Mixture Bayes intervals, and can be obtained by specifying the prior distribution for $\delta$
as a zero-mean normal with the prior variance $s_0^2$ such that $\sigma_0^2 = N\times s_0^2$ is very large. 
When P-values are selected based on a cut-off value or their magnitude, P-intervals can still be a poor approximation to a distribution with $\sigma^2_0$ as large as 10. For example, the last row of Table \ref{tab:ZminP} demonstrates that P-intervals are still biased for $\sigma^2_0=10$ in terms of the coverage when constructed for the minimum P-value taken from multiple-testing experiments with 10,000 tests. Multiple-testing on the scale of genome-wide studies would further degrade the coverage of P-intervals. Using the commonly used asymptotically normal statistic for odds ratio as an example, we emphasize that the standardized values $\delta$ can be bounded, and in this case, the bounds ($L$,$U$) are approximately $-0.66 < \delta < 0.66$. This places specific restrictions on how large $s_0^2$ can be. For the zero-mean normal prior, $s_0$=0.66/3 is still unreasonably large, and in general, even for prior distributions concentrated at these bounds, $s_0^2 \le (U-L)^2/4$ by Popoviciu's inequality.

\section*{Conflict of Interest}

The authors declare no conflict of interest.

\section*{Acknowledgements}
This research was supported in part by the Intramural Research Program of the NIH, National Institute of Environmental Health Sciences.

\clearpage
\begin{flushleft}
\bibliography{pintervals}
\end{flushleft}
\clearpage
\section*{Tables}
\begin{table}[H]
\centering
\begin{tabular}{lcccc}
  \hline
   Type of P-value selection & Prior variance ($\sigma_0^2$) & Conjugate Bayes  & Mixture Bayes & P-interval \\ 
  \hline
  0 $\leq \text{P-value} \leq$ 1 & 0.25 & 80.1\% & 80.2\% & 80.2\% \\ 
  (no selection) & 0.50 & 80.0\% & 80.0\% & 79.9\% \\ 
                  & 1.00 & 80.0\% & 80.0\% & 80.0\% \\ 
                  & 3.00 & 80.4\% & 80.4\% & 80.4\% \\ 
                  & 5.00 & 80.2\% & 80.2\% & 80.3\% \\ 
                  & 10.00 & 80.1\% & 80.1\% & 80.1\% \\ 
  \hline
  0.045 $\leq \text{P-value} \leq$ 0.055 & 0.25 & 79.8\% & 79.8\% & 58.4\% \\ 
                  & 0.50 & 80.1\% & 80.1\% & 66.7\% \\ 
                  & 1.00 & 79.8\% & 79.8\% & 73.5\% \\ 
                  & 3.00 & 80.0\% & 80.0\% & 80.2\% \\ 
                  & 5.00 & 79.9\% & 79.9\% & 80.7\% \\ 
                  & 10.00 & 80.1\% & 80.1\% & 80.8\% \\ 
  \hline
  0 $\leq \text{P-value} \leq$ 0.05 & 0.25 & 80.0\% & 80.0\% & 46.0\% \\ 
                  & 0.50 & 80.1\% & 80.1\% & 55.4\% \\ 
                  & 1.00 & 80.2\% & 80.2\% & 65.5\% \\ 
                  & 3.00 & 79.8\% & 79.8\% & 75.7\% \\ 
                  & 5.00 & 80.4\% & 80.4\% & 78.4\% \\ 
                  & 10.00 & 80.3\% & 80.3\% & 79.5\% \\ 
  \hline
  0 $\leq \text{P-value} \leq$ 0.001 & 0.25 & 80.1\% & 80.1\% & 17.0\% \\ 
                  & 0.50 & 80.1\% & 80.1\% & 29.7\% \\ 
                  & 1.00 & 80.0\% & 79.9\% & 47.6\% \\ 
                  & 3.00 & 80.0\% & 80.0\% & 70.2\% \\ 
                  & 5.00 & 79.9\% & 79.9\% & 75.4\% \\ 
                  & 10.00 & 79.7\% & 79.8\% & 78.2\% \\ 
  \hline
  $5\times 10^{-8}\leq \text{P-value} \leq 1.5 \times 10^{-7}$ & 3.00 & 80.1\% & 80.1\% & 62.8\% \\ 
                  & 5.00 & 79.5\% & 79.5\% & 72.6\% \\ 
                  & 10.00 & 79.8\% & 79.8\% & 78.3\% \\ 
  \hline 
  $5 \times 10^{-9} \leq \text{P-value} \leq 1.5 \times 10^{-8}$ & 3.00 & 80.0\% & 80.0\% & 60.6\% \\ 
                  & 5.00 & 79.9\% & 80.0\% & 71.8\% \\ 
                  & 10.00 & 80.2\% & 80.2\% & 78.1\% \\ 
   \hline
   
\end{tabular}
\caption{The empirical coverage probabilities of the 80\% prediction intervals for a two-sample Z-test. The table illustrates the effect of the observed P-value thresholding, e.g., selection of statistically significant P-values at 5\% level, on the empirical coverage.    
}
\label{tab:Zselection}
\end{table}

\begin{table}[H]
\centering
\begin{tabular}{lcccc}
  \hline
   Number of tests & Prior variance ($\sigma_0^2$) & Conjugate Bayes  & Mixture Bayes & P-interval \\ 
  \hline
  $L=10$ & 0.25 & 80.4\% & 80.4\% & 63.8\% \\ 
  & 0.50 & 79.9\% & 79.9\% & 66.2\% \\ 
  & 1.00 & 80.6\% & 80.6\% & 70.4\% \\ 
  & 3.00 & 80.0\% & 80.0\% & 75.1\% \\ 
  & 5.00 & 80.1\% & 80.1\% & 76.7\% \\ 
  & 10.00 & 80.1\% & 80.1\% & 78.3\% \\ 
  \hline
  $L=100$ & 0.25 & 79.8\% & 79.8\% & 35.7\% \\ 
  & 0.50 & 80.2\% & 80.2\% & 42.3\% \\ 
  & 1.00 & 79.9\% & 79.9\% & 51.1\% \\ 
  & 3.00 & 79.6\% & 79.6\% & 65.1\% \\ 
  & 5.00 & 80.0\% & 80.0\% & 70.0\% \\ 
  & 10.00 & 79.8\% & 79.8\% & 74.5\% \\ 
  \hline
  $L=1,000$ & 0.25 & 80.0\% & 80.1\% & 16.9\% \\ 
  & 0.50 & 79.9\% & 79.8\% & 23.9\% \\ 
  & 1.00 & 80.0\% & 79.9\% & 35.0\% \\ 
  & 3.00 & 80.0\% & 79.9\% & 55.5\% \\ 
  & 5.00 & 79.7\% & 79.6\% & 63.1\% \\ 
  & 10.00 & 80.2\% & 80.1\% & 70.7\% \\ 
  \hline
  $L=10,000$ & 0.25 & 80.1\% & 80.1\% & 07.2\% \\ 
  & 0.50 & 80.1\% & 80.0\% & 12.9\% \\ 
  & 1.00 & 79.8\% & 79.6\% & 23.1\% \\ 
  & 3.00 & 79.7\% & 79.1\% & 46.2\% \\ 
  & 5.00 & 80.2\% & 79.6\% & 56.5\% \\ 
  & 10.00 & 80.2\% & 79.5\% & 66.5\% \\ 
   \hline
\end{tabular}
\caption{The empirical coverage probabilities of the 80\% prediction intervals for a two-sample Z-test. The table illustrates the effect of selecting the most significant P-value (out of $L$ test) on the P-interval coverage.} 
\label{tab:ZminP}
\end{table}

\begin{table}[H]
\centering
\begin{tabular}{lccccc}
  \hline
   Number of tests  & Prior variance ($\sigma_0^2$) & Bayes ($\sigma_0^2/2$) & Bayes ($2 \sigma_0^2)$ & P-interval \\ 
  \hline
  $L=1$     & 0.5 & 77.5\% & 81.7\% & 80.3\% \\
            & 1   & 76.5\% & 81.5\% & 79.7\% \\
  \hline
  $L=10$     & 0.25 & 77.6\% & 81.1\% & 63.8\% \\
             & 0.5  & 75.9\% & 80.8\% & 66.2\% \\
   \hline
  $L=100$    & 3    & 70.4\% & 77.8\% & 65.1\% \\
             & 1    & 72.1\% & 77.2\% & 51.1\% \\
   \hline
  $L=1000$   & 0.25 & 76.4\% & 78.2\% & 16.9\% \\
             & 0.5  & 72.9\% & 75.8\% & 23.9\% \\
   \hline
  $L=10\;000$  & 3    & 62.0\% & 72.9\% & 46.2\% \\
             & 10   & 69.5\% & 77.1\% & 66.5\% \\
   \hline
\end{tabular}
\caption{The effect of the prior variance misspecification on the coverage of the Bayesian-type prediction interval.} 
\label{Tab:bias}
\end{table}

\begin{table}
  \centering
  \resizebox{1\textwidth}{!}{
    \begin{tabular}[t!]{llccccccc} \hline
      SNP & Disorder & Cases & Controls & One-sided P-value & \multicolumn{4}{c}{Prediction intervals, $-\log_{10}\; \text{P}$}\\ \cline{6-9}
          &&&&& Conjugate Bayes$^{\dagger}$  & Mixture Bayes$^{\dagger}$ & Mixture Bayes$^{\ddagger}$ & Lazerroni et al.\\ \hline
      rs2535629 & ADHD & 2787 & 2635 & 0.1005 & (0.01, 1.63) & (0.01, 1.63) & (0.01, 1.64) & (0.03, 4.59)\\
          & ASD & 4949 & 5314 & 0.098 & (0.01, 1.66) & (0.01, 1.66) & (0.01, 1.66) & (0.03, 4.62) \\ 
          & BP & 6990 & 4820 & 3.305e-06 & (0.01, 1.77) & (0.01, 1.77) & (0.03, 6.16) & (1.38, 12.77) \\
          & MDD & 9227 & 7383 & 0.000108 & (0.01, 1.80) & (0.01, 1.80) & (0.02, 5.28) & (0.75, 10.31)  \\
          & Schizophrenia & 9379 & 7736 & 3.355e-05&   (0.01, 1.82) &  (0.01, 1.82) & (0.02, 6.40) & (0.95, 11.16)\\
          & All & 33 332 & 27 888 & 1.27e-12 & (0.06, 2.93) & (0.06, 2.91) & (4.83, 17.65) & (4.93, 22.13) \\ \hline
     \end{tabular}}
  \caption{Revised predictions based on recent results from the Psychiatric Genomics Consortium\cite{lazzeroni2014p} with the prior effect size distribution estimated for the bipolar disorder susceptibility loci.\cite{chen2013genome} Abbreviations: ADHD, attention deficit-hyperactivity disorder; ASD, autism spectrum disorder; BP, bipolar disorder; MDD, major depressive disorder. $^{\dagger}$The prior effect size distribution using the conjugate model with the variance estimated based on the tabulated values of effect sizes reported in Chen et al. \cite{chen2013genome} $^{\ddagger}$The prior effect size distribution specified directly by the estimates reported in Chen et al. \cite{chen2013genome}  }
  \label{tab:bp}
\end{table}

\begin{table}
  \centering
  \resizebox{1\textwidth}{!}{
    \begin{tabular}[t!]{llccccccc} \hline
      SNP & Disorder & Cases & Controls & One-sided P-value & \multicolumn{4}{c}{Prediction intervals, $-\log_{10}\; \text{P}$}\\ \cline{6-9}
          &&&&& Conjugate Bayes$^{\dagger}$  & Mixture Bayes$^{\dagger}$ & Mixture Bayes$^{\ddagger}$ & Lazerroni et al.\\ \hline
      rs2535629 & ADHD & 2787 & 2635 & 0.1005 & (0.01, 1.63) & (0.01, 1.63) & (0.01, 1.63) & (0.03, 4.59)\\
          & ASD & 4949 & 5314 & 0.098 & (0.01, 1.66) & (0.01, 1.66) & (0.01, 1.66) & (0.03, 4.62) \\ 
          & BP & 6990 & 4820 & 3.305e-06 & (0.01, 1.76) & (0.01, 1.76) & (0.03, 8.23) & (1.38, 12.77) \\
          & MDD & 9227 & 7383 & 0.000108 & (0.01, 1.80) & (0.01, 1.80) & (0.01, 4.13) & (0.75, 10.31)  \\
          & Schizophrenia & 9379 & 7736 & 3.355e-05&   (0.01, 1.82) &  (0.01, 1.82) & (0.02, 6.03) & (0.95, 11.16)\\
          & All & 33 332 & 27 888 & 1.27e-12 & (0.06, 2.89) & (0.06, 2.89) & (4.37, 19.05) & (4.93, 22.13) \\ \hline
     \end{tabular}}
  \caption{Revised predictions based on recent results from the Psychiatric Genomics Consortium\cite{lazzeroni2014p} with the prior effect size distribution estimated for cancer risk loci.\cite{park2010estimation} Abbreviations: ADHD, attention deficit-hyperactivity disorder; ASD, autism spectrum disorder; BP, bipolar disorder; MDD, major depressive disorder. $^{\dagger}$The prior effect size distribution using the conjugate model with the variance estimated based on the tabulated values of effect sizes reported in Park et al. \cite{park2010estimation} $^{\ddagger}$The prior effect size distribution specified directly by the estimates reported in Park et al. \cite{park2010estimation}  }
  \label{tab:cancer}
\end{table}

\clearpage
\section*{Supplementary Information}
\subsection*{Simulations} A statement where $P_\text{obt}$ is given any value, such as 0.05, is a type of selection that induces selection bias, commonly described as the winner's curse. Our simulations are designed to demonstrate that sort of bias for P-intervals. On the other hand, when we are equipped with knowledge about the actual underlying effect size distribution, the resulting intervals are expected to be immune to selection bias. The most straightforward scenario is to restrict computation of the intervals to P-values in a narrow interval around a value, such as 0.05, and see empirically what the actual coverage is, compared to the declared 80\%. We proceeded with this scenario as follows. Let the prior distribution for the mean of a Z-statistic be $\mu \sim \sqrt{N} \Phi(m_0, s_0^2)$. In our simulations we set $m_0 = 0$. Simulation steps are as follows:
\begin{enumerate}
\item Draw a value of $\mu$ from its assumed distribution. Simulate data by taking two samples of normal observations with the population means 0 and $\mu$ and compute a Z-statistic, $z_\text{obt}$  (this step can be simplified by drawing $z_\text{obt}$ directly from $\Phi(\mu, 1)$). Calculate the P-value,  $P_\text{obt}$, from $z_\text{obt}$. If $P_\text{obt}$ does not fall within a specified range, e.g., 0.045 to 0.055, discard $z_\text{obt}$. Repeat this step until $P_\text{obt}$ falls within the predefined range.
\item Simulate data as in the previous step or draw $z_\text{rep}$ directly from $\Phi(\mu, 1)$.
\item Calculate the P-interval using the prediction distribution $\Phi(z_\text{obt}, 2)$. Check whether the replication value falls within the interval. 
\item Calculate Bayesian prediction intervals and check whether the replication value falls within these interval.
\end{enumerate}

Next, we repeat the above steps 50,000 times. At the end, we calculate the proportion of times when the replication value was within the studied intervals.

We studied two additional types of P-value selection: (i) inclusion of P-values that are smaller than a predefined threshold, that is, we kept only those P-values that are less than some value, e.g. $P_\text{obt} \le 0.05$; (ii) selection of the minimum P-value from a multiple testing experiment with $L$ tests (in this modification we draw $L$ statistics $z_\text{obt}$ and keep the maximum one that corresponds to the smallest P-value).

\subsection*{Supplementary tables}
\renewcommand{\thetable}{S\arabic{table}}
\setcounter{table}{0}

 \begin{table}[H]
 \centering
 \begin{tabular}{lcc}
   \hline
   Type of P-value selection & Prior variance ($\sigma_0^2$) & Mixture Bayes coverage\\ 
   \hline
   0 $\leq \text{P-value} \leq$ 1 & 0.25 & 79.7\% \\ 
   (no selection)            & 0.50 & 79.8\% \\ 
                       & 1.00 & 80.1\% \\ 
                       & 3.00 & 80.1\% \\ 
                       & 5.00 & 79.9\% \\ 
                       & 10.00 & 79.9\% \\ 
   \hline
   0.045 $\leq \text{P-value} \leq$ 0.055 & 0.25 & 79.4\% \\ 
                       & 0.50 & 79.7\% \\ 
                       & 1.00 & 80.5\% \\ 
                       & 3.00 & 80.5\% \\ 
                       & 5.00 & 80.7\% \\ 
                       & 10.00 & 79.5\% \\ 
   \hline
   0 $\leq \text{P-value} \leq$ 0.05 & 0.25 & 79.9\% \\ 
                       & 0.50 & 79.7\% \\ 
                       & 1.00 & 80.7\% \\ 
                       & 3.00 & 79.7\% \\ 
                       & 5.00 & 80.0\% \\ 
                       & 10.00 & 79.9\% \\ 
   \hline
   0 $\leq \text{P-value} \leq$ 0.001 & 0.25 & 79.9\% \\ 
                       & 0.50 & 79.0\% \\ 
                       & 1.00 & 80.2\% \\ 
                       & 3.00 & 80.5\% \\ 
                       & 5.00 & 79.9\% \\ 
                       & 10.00 & 79.6\% \\ 
 
   \hline
 \end{tabular}
 \caption{The empirical coverage probabilities of the 80\% mixture-Bayes prediction intervals for a two-sample $t$-test under selection of P-values.} 
\label{tab:si1}
\end{table}

 \begin{table}[H]
 \centering
 \begin{tabular}{lcc}
   \hline
   Number of tests & Prior variance ($\sigma_0^2$) & Mixture Bayes coverage \\ 
   \hline
   $L=10$ & 0.25 & 79.2\% \\ 
                  & 0.50 & 79.9\% \\ 
                  & 1.00 & 80.3\% \\ 
                  & 3.00 & 79.8\% \\ 
                  & 5.00 & 80.1\% \\ 
                  & 10.00 & 79.3\% \\ 
   \hline
   $L=100$ & 0.25 & 80.3\% \\ 
                  & 0.50 & 79.9\% \\ 
                  & 1.00 & 79.9\% \\ 
                  & 3.00 & 80.7\% \\ 
                  & 5.00 & 80.1\% \\ 
                  & 10.00 & 80.4\% \\ 
   \hline
   $L=1000$ & 0.25 & 79.1\% \\ 
                  & 0.50 & 80.0\% \\ 
                  & 1.00 & 80.4\% \\ 
                  & 3.00 & 80.1\% \\ 
                  & 5.00 & 79.9\% \\ 
                  & 10.00 & 79.9\% \\ 
   \hline
   $L=10\;000$ & 0.25 & 79.6\% \\ 
                  & 0.50 & 80.2\% \\ 
                  & 1.00 & 79.2\% \\ 
                  & 3.00 & 79.8\% \\ 
                  & 5.00 & 79.9\% \\ 
                  & 10.00 & 79.3\% \\ 
   \hline
 \end{tabular}
 \caption{The empirical coverage probabilities of the 80\% mixture-Bayes prediction intervals constructed for the most significant results out of $L$ two-sample $t$-tests.} 
\label{tab:si2}
\end{table}

\end{document}